 \documentclass[preprint,12pt]{elsarticle}

\usepackage{amsmath}
\usepackage{amssymb}
\usepackage[ruled,vlined]{algorithm2e}
\SetKwBlock{Repeat}{repeat}{}
\usepackage{gensymb}
\usepackage{setspace}
\usepackage{graphics}
\usepackage{pdflscape}
\usepackage{afterpage}
\usepackage{capt-of}
\usepackage[usenames,dvipsnames,svgnames,table]{xcolor}

\PassOptionsToPackage{linktocpage}{hyperref}

\newcommand{\algorithmfootnote}[2][\footnotesize]{%
  \let\old@algocf@finish\@algocf@finish
  \def\@algocf@finish{\old@algocf@finish
    \leavevmode\rlap{\begin{minipage}{\linewidth}
    #1#2
    \end{minipage}}%
  }%
}

\journal{Nuclear Physics B}

\begin{document}

\begin{frontmatter}

\title{Data-driven control of micro-climate in buildings: an event-triggered reinforcement learning approach}

\author[labela]{Ashkan Haji Hosseinloo\corref{cor1}}
\ead{ashkanhh@mit.edu}
\author[labelc]{Alexander Ryzhov}
\author[labelc]{Aldo Bischi}
\author[labelc]{Henni Ouerdane}
\author[labelb]{Konstantin Turitsyn}
\author[labela]{Munther A. Dahleh}

\address[labela]{Laboratory for Information and Decision Systems, MIT, USA}
\address[labelb]{D. E. Shaw Group, New York, NY 10036 USA}
\address[labelc]{Center for Energy Science and Technology, Skolkovo Institute of Science and Technology, 3 Nobel Street, Skolkovo, Moscow Region 121205, Russia}






\begin{abstract}
{ \color{black}
Smart buildings have great potential for shaping an energy-efficient, sustainable, and more economic future for our planet as buildings account for approximately 40\% of the global energy consumption. Future of the smart buildings lies in using sensory data for adaptive decision making and control that is currently gloomed by the key challenge of learning a good control policy in a short period of time in an online and continuing fashion. To tackle this challenge, an event-triggered -- as opposed to classic time-triggered -- paradigm, is proposed in which learning and control decisions are made when events occur and enough information is collected. Events are characterized by certain design conditions and they occur when the conditions are met, for instance, when a certain state threshold is reached. By systematically adjusting the time of learning and control decisions, the proposed framework can potentially reduce the variance in learning, and consequently, improve the control process. We formulate the micro-climate control problem based on semi-Markov decision processes that allow for variable-time state transitions and decision making. Using extended policy gradient theorems and temporal difference methods in a reinforcement learning set-up, we propose two learning algorithms for event-triggered control of micro-climate in buildings. We show the efficacy of our proposed approach via designing a smart learning thermostat that simultaneously optimizes energy consumption and occupants' comfort in a test building.
}
\end{abstract}


\end{frontmatter}


\section{Introduction}
\label{intro}
Buildings account for approximately 40\% of global energy consumption about half of which is used by heating, ventilation, and air conditioning (HVAC) systems \cite{nejat2015global, wei2017deep}, the primary means to control micro-climate in buildings. Furthermore, buildings are responsible for one-third of global energy-related greenhouse gas emissions \cite{nejat2015global}. Hence, even an incremental improvement in the energy efficiency of buildings and  HVAC systems goes a long way towards building a sustainable, more economic, and energy-efficient future. In addition to their economic and environmental impacts, HVAC systems can also affect productivity and decision-making performance of occupants in buildings through controlling indoor thermal and air quality \cite{satish2012co2, wargocki2017ten}. For all these reasons  micro-climate control in buildings is an important issue for its large-scale economic, environmental, and health-related and societal effects. \\

The main goal of the micro-climate control in buildings is to minimize the building's (mainly HVAC's) energy consumption while improving or respecting some notion of occupants' comfort. Despite its immense importance, micro-climate control in buildings is often very energy-inefficient. HVAC systems are traditionally controlled by rule-based strategies and heuristics where an expert uses best practices to create a set of rules that control different HVAC components such as rule-based \texttt{ON/OFF} and conventional PID controllers \cite{levermore2013building, dounis2009advanced}. These control methods are often far from optimal as they do not take into account the building thermodynamics and stochasticities such as weather conditions or occupancy status. To overcome some of these shortcomings, more advanced model-based approaches have been proposed. In this category, model predictive control (MPC) is perhaps the most promising and extensively-studied method in the context of buildings climate control \cite{oldewurtel2012use,ryzhov2019model,afram2014theory, smarra2018data}.\\

\textcolor{black} {Despite its potential benefits, performance and reliability of MPC and other model-based control methods depend highly on the accuracy of the building thermodynamics model and prediction of the stochastic disturbances. However, developing an accurate model for a building is extremely time-consuming and resource-intensive, and hence, not practical in most cases. Moreover, a once accurately developed model of a building could become fairly inaccurate over time due to, for instance, renovation or wear and tear of the building. Furthermore, at large scales,  MPC like many other advanced model-based techniques may require formidable computational power if a real-time (or near real-time) solution is required \cite{marantos2019rapid}. Last but not least, traditional and model-based techniques are inherently building-specific and not easily transferable to other buildings.}\\

To remedy the above-mentioned issues of the model-based climate control in buildings and towards building \textit{smart} homes, data-driven approaches for HVAC control have attracted much interest in the recent years. The concept of \textit{smart} homes where household devices (e.g., appliances, thermostats, and lights) can operate efficiently in an autonomous, coordinated, and adaptive fashion, has been around for a couple of decades \cite{mozer1998neural}. However, with recent advances in Internet of Things (IoT) technology (cheap sensors, efficient data storage, etc.) on the one hand \cite{minoli2017iot}, and immense progress in data science and machine learning tools on the other hand, the idea of smart homes with data-driven HVAC control systems looks ever more realistic.\\

Among different data-driven control approaches, reinforcement learning (RL) has found more attention in the recent years due to recent algorithmic advances in this field as well as its ability to learn efficient control policies solely from experiential data via trial and error. This study focuses on an RL approach and hence, we next discuss some of the related studies using reinforcement learning for energy-efficient controls in buildings followed by our contribution.\\

The remaining of this article is organized as follows. Section \ref{related} reviews the related literature, discusses their limitations, and highlights our contributions. The micro-climate control problem is stated and mathematically formulated in section \ref{problemstatement} where the idea of the switching manifolds is also introduced. Section \ref{SMDP} introduces preliminaries of a semi-Markov decision process (SMDP), and then, delineates how the original control problem is formulated in the SMDP framework. Learning-based control algorithms are presented in section \ref{RL}. These algorithms are implemented via simulation on two different building models and the results are presented and discussed in section \ref{results}. Finally, section \ref{conclusion} concludes the paper with a summary and some open research problems and future work.
\section{Related work and contributions}
\label{related}
\subsection{Tabular RL}
The Neural Network House project \cite{mozer1998neural} is perhaps the first application of reinforcement learning in building energy management system. In this seminal work, the author explains how tabular Q-learning, one of the early versions of the popular Q-learning approach in RL, was employed to control lighting in a residential house so as to minimize energy consumption subject to occupants' comfort constraint \cite{mozer1997parsing}. Tabular Q-learning was later used in a few other studies for controlling passive and active thermal storage inventory in commercial buildings \cite{liu2006experimental1, liu2006experimental2}, heating system\cite{barrett2015autonomous}, air-conditioning and natural ventilation through windows \cite{chen2018optimal}, photovoltaic arrays and geothermal heat pumps \cite{yang2015reinforcement}, and lighting and blinds \cite{cheng2016satisfaction}.\\

Given fully observable state and infinite exploration, tabular Q-learning is guaranteed to converge on an optimal policy. However, the tabular version of Q-learning is limited to systems with discrete states and actions, and becomes very data-intensive, hence very slow at learning, when the system has a large number of state-action combinations. For instance, the simulated RL training in \cite{liu2006experimental2} for a fairly simple building required up to 6000 days (roughly 17 years) of data collection. To remedy some of these issues, other versions of Q-learning such as Neural Fitted Q-Iteration (NFQ) and deep RL (DRL) were employed where function approximation techniques are used to learn an approximate function of the true action-value function, aka the Q-function.

\subsection{RL with action-value function approximation}
Dalamagkidis et al. \cite{dalamagkidis2007reinforcement} used a linear function approximation technique to approximate the Q-function in their Q-learning RL to control a heat pump and an air ventilation subsystem using sensory data on indoor and outdoor air temperature, relative humidity, and $\mathrm{CO_2}$ concentration. Fitted Q-Iteration (FQI) developed by Ernst et al. \cite{ernst2005tree} is a batch RL method that iteratively estimates the Q-function given a fixed batch of past interactions. In a series of studies \cite{ruelens2015learning, ruelens2016residential, ruelens2016reinforcement}, Ruelens et al. studied the application of FQI batch RL to schedule thermostatically-controlled HVAC systems, such as heat pumps and electric water heaters, in different demand-response set-ups. An online version of FQI that uses a neural network, Neural Fitted Q-Iteration, was proposed by \cite{riedmiller2005neural}. Marantos et al. \cite{marantos2018towards} applied NFQ batch RL to control the thermostat set-point of a single-zone building where input state was four-dimensional (outdoor and indoor temperatures, solar radiance, and indoor humidity) and action was one-dimensional with three discrete values.\\

Immense algorithmic and computational advancements in deep neural networks in the recent years have given rise to the field of deep reinforcement learning where deep neural networks are employed often for function approximation. This has resulted in numerous DRL algorithms (DQN, DDQN, RBW, A3C, DDPG, etc.) in the past few years, some of which have been employed for data-driven micro-climate control in buildings. Wei et al. \cite{wei2017deep} claim to be the first to apply DRL to HVAC control problem. They used Deep Q-Network (DQN) algorithm \cite{mnih2015human} to approximate the Q-function with discrete number of actions. To remedy some of the issues of the DQN algorithm such as overestimation of action values, improvements to this algorithm have been made, resulting in a bunch of other algorithms, such as Double DQN (DDQN) \cite{van2016deep} and Rainbow (RWB) \cite{hessel2018rainbow}. Avendano et al. \cite{avendano2018data} applied DDQN and RWB algorithms to optimize energy efficiency and comfort in a two-zone apartment; they considered temperature and $\mathrm{CO_2}$ concentration for comfort and used heating and ventilation costs for energy efficiency.

\subsection{RL with policy function approximation}
All the above-mentioned RL-based studies rely on learning the optimal state-value or action-value functions based on which the optimal policy is derived. Parallel to the value-based approach, there is a policy-based approach where the RL agent tries to directly learn the optimal policy (i.e., the control law). Policy gradient algorithms are perhaps the most popular class of RL algorithms in this approach. The basic idea behind these algorithms is to adjust the parameters of the policy in the direction of a performance gradient \cite{sutton2000policy, silver2014deterministic}. A distinctive advantage of policy gradient algorithms is their ability to handle continuous actions as well as stochastic policies. Wang et al. \cite{wang2017long} employed Monte Carlo actor-critic policy gradient RL with long short-term memory (LSTM) actor and critic networks to control HVAC system of a single-zone office. Deep Deterministic Policy Gradient (DDPG) algorithm \cite{lillicrap2015continuous} is another powerful algorithm in this class that handles deterministic policies. DDPG was used in \cite{gao2019energy} and \cite{li2019transforming} to control energy consumption in a single-zone laboratory and two-zone data center buildings, respectively.

\subsection{Limitations of RL and its application to micro-climate control}
\label{limitations}
Despite the recent advances in RL, sample efficiency is still the bottleneck for many real-world applications with slow dynamics. Building micro-climate control is one such application since thermodynamics in buildings (e.g., change in building's temperature or humidity) is a relatively slow process. The time-intensive process of data collection makes the online training of the RL algorithms so long that it practically becomes impossible to have a plug \& play RL-based controller for HVAC systems. For instance, training the DQN RL algorithm in \cite{wei2017deep} for a single-zone building required about 100 months of sensory data. The required data collection time for training the DDQN and RWB algorithms in \cite{avendano2018data} was reported as 120 and 90 months, respectively. A few different techniques have been proposed to alleviate the sample complexity of the RL approach when it comes to real-world applications, in particular buildings, that are discussed next.\\

Multiple time scales in some real-world applications is one reason for the sample inefficiency of many RL algorithms. For instance, for precise control of a set-point temperature it is more efficient to design a controller that works on a coarse time scale in the beginning when the temperature is far from the set-point temperature, and on a finer time scale otherwise. To address this issue, double and multiple scales reinforcement learning are proposed in \cite{riedmiller1998high, li2015multi}. Reducing the system's dimension, if possible, is another way to shorten the online training period. Different dimensionality reduction techniques such as auto-encoder \cite{ruelens2015learning} and Convolutional Neural Networks (CNN) \cite{claessens2016convolutional} were used in RL-based building energy management control where the system states were high dimensional.\\

Another approach to reduce the training period is based on developing a data-driven model first, and then use it for offline RL training or direct planning. This approach is similar to the Dyna architecture \cite{sutton1991dyna, sutton2018reinforcement} and is often referred to as model-based RL \cite{hosseinloo2020event}. Costanzo et al. \cite{costanzo2016experimental} used neural networks to learn temperature dynamics of a building's heating system to feed training of their FQI RL algorithm while Nuag et al. \cite{naug2019online} used support vector regression to develop consumption energy model of a commercial building for training of their DDPG algorithm. In \cite{nagy2018deep} and \cite{kazmi2018gigawatt} data-driven models of thermal systems are developed in the form of neural networks and transition matrix \textcolor{black} {\footnote{\textcolor{black}{A transition matrix describes a probabilistic model of a dynamic system with probabilities of system transitions from one state to another.}}} of partially observable Markov decision processes (POMDPs), respectively, which are then used for finite horizon planning. As another example, Kazmi et al. \cite{kazmi2019multi} used muti-agent RL to learn a probabilistic model of identical thermostatically-controlled loads, which was then used for deriving the optimal policy by Monte Carlo techniques.\\

In addition to the issue of large sample complexity that is inherent to most RL algorithms, there are issues on how the RL techniques are employed for the micro-climate control problem. Similar to many other studies about RL applications to physical sciences, there are two main issues with the above-reviewed studies; first, they model and solve the micro-climate control problem as an \textit{episodic-task} problem with \textit{discounted reward} while it should be modeled as a \textit{continuing-task} problem with \textit{average reward}. Average reward is really what matters in continuing-task problems and greedily maximizing discounted future values does not necessarily maximize the average reward \cite{naik2019discounted}. In particular, solutions that fundamentally rely on episodes are likely to fare worse than those that fully embrace the continuing task setting.\\

Second, in all these studies, the control problem is modeled based on Markov decision processes (MDPs) where learning and decision-making occur at fixed sampling rate. The fixed time intervals between decisions (control actions) is restrictive in continuous-time problems; a large interval (low sampling rate) deteriorates the control accuracy while a small interval (high sampling rate) could drastically affect the learning quality. For instance, as reported in \cite{munos2006policy} among others, policy gradient estimate is subject to variance explosion when the discretization time-step tends to zero. The intuitive reason for that problem lies in the fact that the number of decisions before getting a meaningful reward grows to infinity. Furthermore, the classic time-triggered learning and control (i.e., learning and control at fixed time intervals) may not be desired in large-scale resource-constrained wireless embedded control systems \cite{heemels2012introduction}.

\subsection{Contributions}
Towards designing plug \& play learning-based controllers for smart buildings, we eliminate the major drawbacks of the RL-based controllers discussed above by proposing an event-triggered learning-based controller. Unlike the conventional periodic paradigm in RL and controls where learning and control take place at periodic times, our proposed controller learns and takes actions aperiodically and when needed. In a nutshell, the major contributions of this paper are as follows:

\begin{itemize}
  \item We formulate the micro-climate control problem as a continuing-task problem with infinite-horizon (undiscounted) average-reward objective;
  \item We introduce the idea of \textit{event-triggered} paradigm along with the notion of \textit{switching manifolds} for data-efficient learning and control with application to HVAC systems;
  \item We formulate the event-triggered control problem in SMDP framework with variable transition times;
  \item We present two event-triggered learning algorithms with application to online micro-climate control in buildings;
  \item We demonstrate the effectiveness of our proposed approach on a small-scale building via simulation in EnergyPlus software.
\end{itemize}

\section{Problem statement}
\label{problemstatement}
\subsection{System dynamics and optimization objective}
\label{systemdynamics}
\textcolor{black} {The aim of this study is to provide a plug \& play control algorithm that can efficiently learn to optimize HVAC energy consumption and occupants' comfort in buildings, with no knowledge of the building's model.} With no loss of generality, we consider a single-zone building with \texttt{ON/OFF} heating system; indeed the methods and concepts that we present in this paper are applicable to more general settings. The building dynamics evolve as:
\begin{equation}
\frac{dT}{dt}=f(T,T_o,u),
\label{Eq:dynamics}
\end{equation}
where, $T(t) \in \mathbb{R}$ represents the building temperature, $T_o(t) \in \mathbb{R}$ is the outdoor temperature (exogenous state), and $u(t) \in \{0,1\}$ denotes the control signal determining the heater's \texttt{ON/OFF} status; $u(t)=1$ switches the heater \texttt{ON} and $u(t)=0$ switches it \texttt{OFF}. The thermal dynamics of the system are characterized by the function $f(.)$ which is \textcolor{black} {an unknown} nonlinear function. Via the control action $u(t)$ we would like to maximize the performance measure $J$, defined as:
\begin{equation}
J = \lim_{T\to \infty} \frac{1}{T} \mathbb{E}\left[\int_0^T\{r_e u(t) + r_c(T-T_d)^2 + r_{sw}\delta(t-t_{sw})\}\,dt \right],
\label{Eq:performance}
\end{equation}
where, $t_{sw}$ is the time when the controller switches from 0 to 1 (the heater switches from \texttt{OFF} to \texttt{ON}) or vice versa, and $\delta(.)$ is the Dirac delta function. The first term of the integrand penalizes the energy consumption while the second and the third terms correspond to occupants' comfort. Specifically, the second term penalizes temperature deviations from a desired set-point temperature ($T_d$) while the third term prevents frequent \texttt{ON/OFF} switching that can consequently reduce the switching noise as well as wear and tear of the heater. The relative effects of these terms are balanced by their corresponding weights, i.e., $r_e$, $r_c$, and $r_{sw}$.

\subsection{Switching manifolds and event-triggered control}
\label{control}
\begin{figure}
 \centering
 \includegraphics[width =.8 \linewidth]{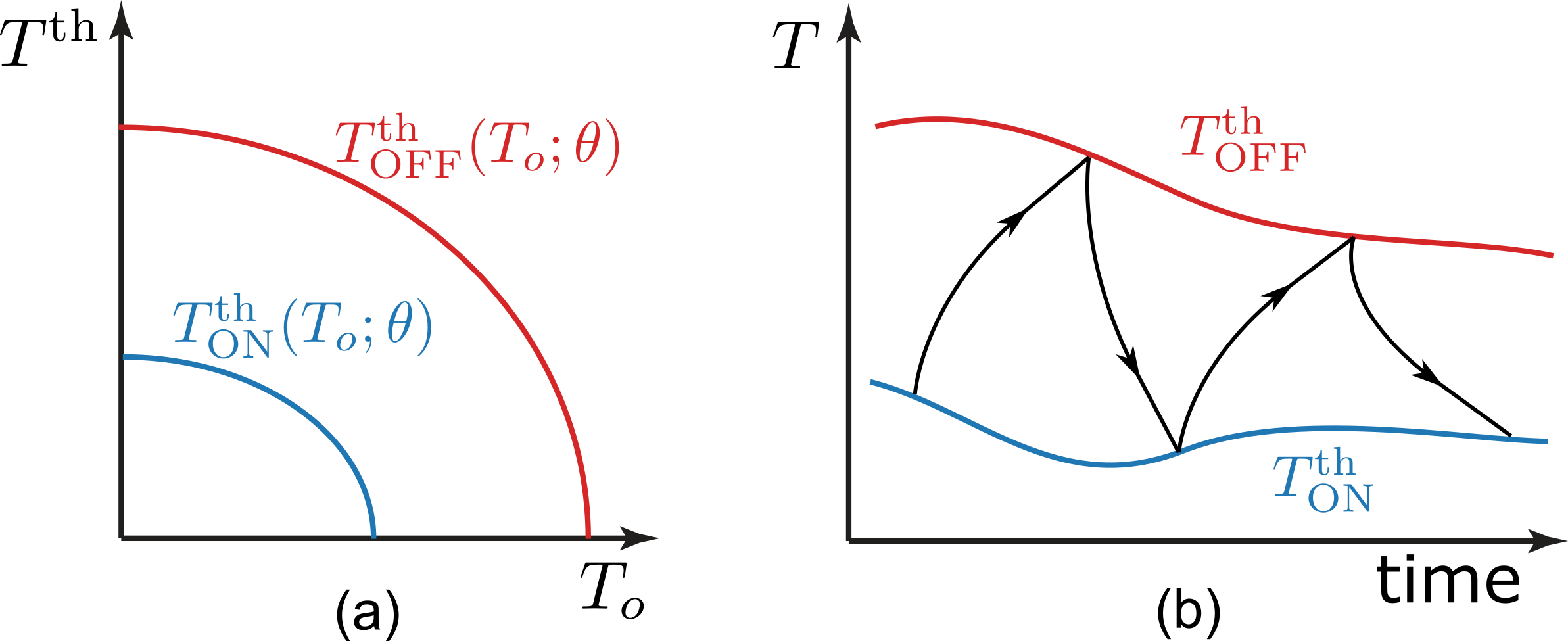}
 \caption{(a) Schematic temperature threshold policy for the one-zone building example where temperature thresholds are parameterized functions of outdoor temperature (b) schematic evolution of building temperature under the threshold control policy}
 \label{Fig:1}
\end{figure}

To reduce the space of possible control policies, we constrain the optimization to a class of parameterized control policies, specifically to threshold policies. This strategy is particularly beneficial in the RL framework since it can potentially reduce learning sample complexity. We characterize the threshold policies by their characteristic \textit{switching manifolds} that are defined in the state space of the system and determine when the control action switches (e.g., \texttt{ON} $\rightleftarrows$ \texttt{OFF} in this study). The control action switches only when the system's state trajectory hits these manifolds which we refer to as \textit{events}. Projecting these manifolds onto the system's indoor temperature results in temperature threshold policies. Figure \ref{Fig:1} (a) illustrates schematically a temperature threshold policy with switch-\texttt{ON} and switch-\texttt{OFF} thresholds for the single-zone building example while Fig.1(b) depicts the evolution of the building temperature under such controller. We can mathematically formulate the control action as:

\begin{equation}
  u(t) =
    \begin{cases}
      0, & \text{if $T(t) \geq T_{\mathrm{OFF}}^{\mathrm{th}}(T,T_o;\theta)$}\\
      1, & \text{if $T(t) \leq T_{\mathrm{ON}}^{\mathrm{th}}(T,T_o;\theta)$}\\
      u(t^-), & \text{otherwise}
    \end{cases},
    \label{Eq:control}
\end{equation}
where, $T_{\mathrm{OFF}}^{\mathrm{th}}(.;\theta)$ and $T_{\mathrm{ON}}^{\mathrm{th}}(.;\theta)$ are temperature thresholds corresponding to the \texttt{OFF} and \texttt{ON} switching  manifolds, respectively that are parameterized in the span of the states by parameter vector $\theta$. The goal is to find the optimal control policy $u^*(t)$ which is parameterized by the optimal parameter vector $\theta^*$, that maximizes the long-run average reward\footnote{Average reward and performance measure are used interchangeably in this paper.} $J$ defined by Eq.(\ref{Eq:performance}). \textcolor{black} {To find the optimal manifolds, or equivalently the optimal thresholds, we need to calculate $J(\theta)$ or its ascent direction. This is not an easy task with no prior knowledge of the system dynamics. To do this, we resort to learning-based control techniques.}\\

 Let us add the heater status, $h_s \in \{0,1\}$ to the state vector. Dynamics of $h_s$ is straightforward; $h_s=0$ when the heater is off and $h_s=1$ otherwise. Also, it changes value only when the system state trajectory hits a switching manifold. By introducing this new state variable we make the controller $u(t)$ memory-less, simply by replacing the third line of Eq.(\ref{Eq:control}) by $h_s(t)$. In this memory-less controller the temperature thresholds $T_{\mathrm{OFF}}^{\mathrm{th}}$ and $T_{\mathrm{ON}}^{\mathrm{th}}$ can be thought of as higher-level control actions. Executing and updating the parameters of the temperature threshold actions could both take place at fixed time steps; however, we avoid it for the reasons explained in section \ref{limitations}. Instead, we restrict the control execution and its parameter update to times when the events occur -- hence the name \textit{event-triggered} control. In other words, the shape of the switching manifolds are controlled and changed only when the system's state trajectory hits them.\\

\begin{figure}
 \centering
 \includegraphics[width =1.0 \linewidth]{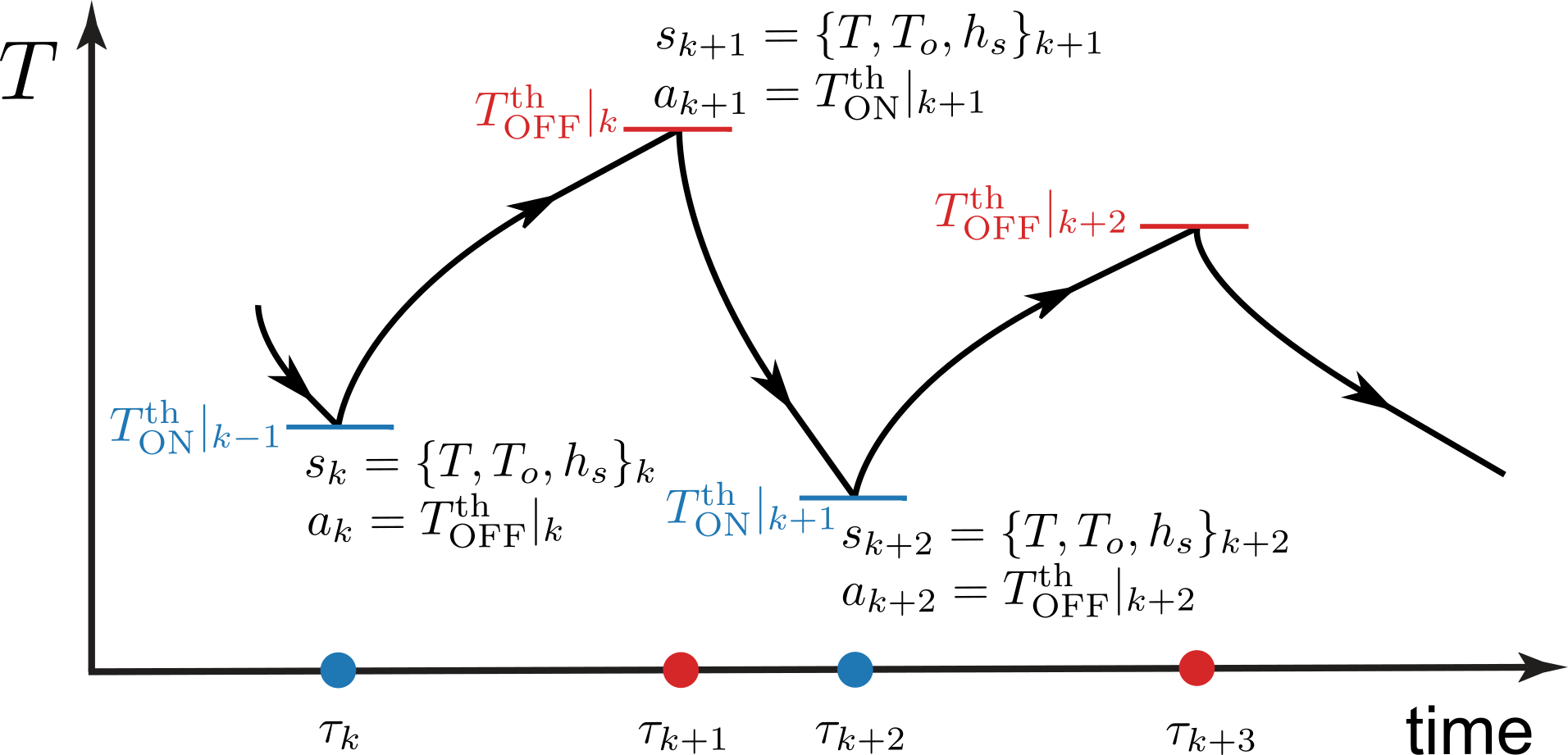}
 \caption{Schematics of building's temperature evolution controlled by threshold policy: temperature thresholds are the control actions resulting in a sequential decision-making process with variable time intervals between decisions. $T^\mathrm{th}_{(.)}|_k$ denotes the temperature threshold action calculated at epoch $\tau_k$.}
 \label{Fig:2}
\end{figure}

Figure \ref{Fig:2} illustrates time history schematic of building's temperature controlled by an event-triggered threshold policy. Given that the control executions and updates occur only at the events, evolution of the system dynamics between the events do not matter so long we can measure the accumulated performance (reward) during this period. As shown in Fig.\ref{Fig:2}, the state vector, $s$, includes the indoor and outdoor temperatures, $T, T_o$, as well as the heater status, $h_s$. At a given event with corresponding state $s_k$ and time $\tau_k$ the controller takes a temperature threshold action, $a_k$, and as a result the system evolves until the next event occurs with a corresponding state $s_{k+1}$ and time of event $\tau_{k+1}$. We refer to the event timestamps, $\tau_k$'s, as epochs. The contribution of this transition to the controller's performance measure can be calculated by Eq.(\ref{Eq:performance}) with the lower and upper bounds of the integral set to $\tau_{k}$ and  $\tau_{k+1}$, respectively.\\

In order to calculate the long-run average reward in this sequential decision-making process, we need transitions' accumulated reward as well as their duration ($\tau_{k+1}-\tau_{k}$). However, the transition period between two consecutive events is not fixed; in fact, it is, in general, a random variable. This is the main reason why the event-triggered control paradigm cannot be formulated based on MDPs. Allowing the state transitions to occur in continuous irregular times makes SMDP framework a better candidate for the event-triggered control problem formulation.

\section{SMDP framework}
\label{SMDP}
In this section we first discuss preliminaries of SMDPs, and then, describe how the SMDP formulation maps to the original micro-climate control problem.
\subsection{Preliminaries}
\label{prelim}
We model the control problem as an SMDP. The main difference between SMDP and MDP is that in SMDP the intervals between decisions are usually random. Put differently, actions can take variable amounts of time to complete. Thus, an SMDP model includes an additional parameter, compared to an MDP model, that defines the duration of an action or the interval between actions.\\

An SMDP can be represented by the five-tuple $(\mathcal{S}, \mathcal{A}, \mathcal{P}, \mathcal{R}, \mathcal{I})$, where $\mathcal{S}$ is the state space, $\mathcal{A}$ is the action space from which the controller (agent) may choose at each decision epoch, and $\mathcal{I}: \mathcal{S} \rightarrow [0,1]$ is the initial state distribution. Different from an MDP model, the transition probability function $\mathcal{P}: \mathcal{S} \times [0,+\infty) \times \mathcal{S} \times \mathcal{A}  \rightarrow [0,1]$ now takes the duration of the actions into account. Let $\tau_k$'s denote the decision epochs with $\tau_0=0$, and $S_k \in \mathcal{S}$ represent the state variable at decision epoch $\tau_k$. Then the function $p(s_{k+1}, t|s_k,a_k)=\mathcal{P}(S_{k+1}=s_{k+1}, \tau_{k+1}-\tau_k \leq t |S_k=s_k,A_k=a_k)$ denotes the probability that action $a_k$ at epoch $\tau_k$ will cause the system to transition from state $s_k$ to state $s_{k+1}$ within $t$ time units\footnote{To simplify notation, we frequently drop the capital-letter random variables in the conditional probabilities.}. The probabilities $p(s_{k+1}, t|s_k,a_k)$ are called semi-Markov kernel. If we let $t\rightarrow +\infty$, the semi-Markov kernel $p(s_{k+1}, +\infty|s_k,a_k)$ will represent the conventional transition probability function of the embedded MDP which we denote it by $p(s_{k+1}|s_k,a_k)$. Also, let $F(t|s_k,a_k)$ denote the probability that the next decision epoch occurs within $t$ time units after the current decision epoch $\tau_k$, given that action $a_k$ is chosen at the current state $s_k$.\\

The reward function of an SMDP, $\mathcal{R}: \mathcal{S} \times \mathcal{A}  \rightarrow \mathbb{R}$, is in general more complex than that of an MDP. Between epochs ($\tau_k\leq t' \leq \tau_{k+1}$) the system evolves based on the so-called \textit{natural process} $W_{t'}$. Let us suppose the reward between two decision epochs consists of two parts; a discrete state-action dependent reward of $g(s_k,a_k)$ and a time-continuous reward accumulated in the transition time at a rate of $c(W_{t'},S_k,A_k)$. We can then write the expected total reward $r(s_k,a_k)$ between the two epochs of $\tau_k$ and $\tau_{k+1}$ as:
\begin{equation}
\begin{split}
r(s_k,a_k) & =  g(s_k,a_k)\\
           &+\mathbb{E}\left[\int_{\tau_k}^{\tau_{k+1}} c(W_{t'},S_k,A_k) dt'|S_k=s_k, A_k=a_k\right].
\end{split}
\label{Eq:reward}
\end{equation}

Let us also define the expected transition time $\tau(s_k,a_k)$, aka dwell or sojourn time, starting at state $s_k$ and under action $a_k$ as:
\begin{equation}
\begin{split}
\tau(s_k,a_k) = & \mathbb{E}\left[\tau_{k+1}-\tau_k|S_k = s_k, A_k=a_k\right]\\
              = & \int_0^\infty tF(dt|s_k,a_k).
\end{split}
\label{Eq:avetime}
\end{equation}

A policy is used to select actions in the SMDP. This policy could be deterministic or stochastic. In a deterministic policy, state space is deterministically mapped into the action space; $a_k=\mu^\theta(s_k)$. However, in a stochastic policy the action $a_k$ is randomly chosen at $s_k$ with the conditional probability density $\pi^\theta(a_k|s_k)$ associated with the policy. We consider parameterized policies where  $\theta \in \mathbb{R}^{d_{\theta}}$ is the parameter vector of dimension $d_{\theta}$. The expected total reward and the expected transition time, defined by Eqs.(\ref{Eq:reward}) and (\ref{Eq:avetime}), can then be written as functions of $\theta$ at each state as:
\begin{equation}
  r(s_k,\theta) =
    \begin{cases}
      r(s_k, \mu^\theta(s_k)), & \text{deterministic policy}\\
      \sum_{a_k}\pi^\theta(a_k|s_k) r(s_k,a_k), & \text{stochastic policy}
    \end{cases},
\end{equation}
and
\begin{equation}
  \tau(s_k,\theta) =
    \begin{cases}
      \tau(s_k, \mu^\theta(s_k)), & \text{deterministic policy}\\
      \sum_{a_k}\pi^\theta(a_k|s_k) \tau(s_k,a_k), & \text{stochastic policy}
    \end{cases}.
\end{equation}

Let us denote the corresponding column vectors of $r(s_k,\theta)$ and $\tau(s_k,\theta)$ by $r(\theta)$ and $\tau(\theta)$, respectively. Assuming ergodicity of the embedded MDP for any $\theta$, let $\rho(\theta)$ designate the row vector steady-state probability distribution of the embedded Markov chain. The infinite-horizon average reward for every initial state $s_0$ of the SMDP is defined as:

\begin{equation}
J(s_0,\theta)  = \lim_{T \to \infty} \frac{1}{T}\mathbb{E}\left[ \left. \int_0^T c(W_{t'},S_{N_{t'}},A_{N_{t'}}) dt' + \sum_{k=0}^{N_{t'}-1} g(S_k,A_k)\; \right| S_0=s_0\right],
\label{Eq:SMDPAveRew1}
\end{equation}
where, $N_{t'}$ denotes the number of decision epochs up to time $t'$. It could be shown that the above average reward is independent of the initial condition under the ergodicity assumption and can be written as \cite{puterman2014markov}:

\begin{equation}
J(\theta)  = \frac{\rho(\theta)r(\theta)}{\rho(\theta)\tau(\theta)}.
\label{Eq:SMDPAveRew2}
\end{equation}

Gradient of the average reward with respect to the policy parameters, i.e., $\nabla_{\theta}J$ plays a key role for improving the policy. It has been shown \cite{li2013basic} that this gradient for stochastic policies takes the form:

\begin{equation}
\nabla_{\theta}J({\theta})= \frac{1}{\rho(\theta)\tau(\theta)} \mathbb{E}_{S_k\sim \rho(\theta), A_k\sim \pi^{\theta}}\left[ \nabla_{\theta}\log \pi^{\theta}(A_k|S_k)Q^{\pi}(S_k,A_k) \right],
\label{Eq:gradstoch}
\end{equation}
where, we refer to $Q^{\pi}(s_k, a_k)$ as differential action-value function and is defined as:
\begin{equation}
Q(s_k, a_k)= r(s_k, a_k) - J(\theta) \tau(s_k, a_k) + \sum_{s_{k+1} \in S} p(s_{k+1}|s_k, a_k) V(s_{k+1}).
\label{Eq:Q}
\end{equation}

The policy-dependent function $V(s_{k+1})$ in Eq.(\ref{Eq:Q}) is referred to as differential state-value function and is defined as:
\begin{equation}
V(s_k) = r(s_k, \theta) - J(\theta) \tau(s_k, \theta) + \sum_{s_{k+1} \in S} \sum_{a_k \in A} p(s_{k+1}|s_k, a_k) V(s_{k+1}).
\label{Eq:V}
\end{equation}

Intuitively, the differential state-value function $V(s_k)$, aka potential function, measures the potential contribution of state $s_k$ to the long-run average reward $J$, for a given policy. Unlike the case of the stochastic policies, the average-reward gradient for deterministic policies has not been studied in the SMDP setup. However, it could be shown \footnote{The formal proof of Eq.(\ref{Eq:graddet}) is beyond the scope of this paper and it will be published soon by the authors in a separate article.} that the gradient for deterministic policies can be written as:
\begin{equation}
\nabla_{\theta}J({\theta}) = \frac{1}{\rho(\theta)\tau(\theta)} \mathbb{E}_{S_k\sim \rho(\theta)}\left[ \nabla_{\theta} \mu^{\theta}(S_k) \left.\nabla_{A_k}Q(S_k,A_k)\right|_{A_k=\mu^{\theta}(S_k)} \right].
\label{Eq:graddet}
\end{equation}

As discussed earlier, the main differences between the average reward setup of SMDP and cumulative discounted reward of MDP are: \textit{average} versus \textit{cumulative} performance measure and \textit{variable} versus \textit{fixed} transition times. Despite these fundamental differences, the average-reward gradients in the SMDP framework provided by Eqs.(\ref{Eq:gradstoch}) and (\ref{Eq:graddet}) look very similar to the gradient of cumulative discounted reward in MDP framework provided by the stochastic and deterministic policy gradient theorems \cite{sutton2000policy, silver2014deterministic}; in fact, all the differences are captured by the notion of \textit{differential} value functions.\\

We can now employ the above gradient formulas to develop sample-based RL algorithms that can improve the policy via, e.g., stochastic gradient ascent optimization. But before delving into the RL algorithms, we further explain how the original control problem in section \ref{problemstatement} maps to the SMDP framework presented in this section.     

\subsection{SMDP formulation of the micro-climate control problem}
\label{mapping}
The micro-climate control problem with threshold policies, as posed in section \ref{problemstatement}, is a sequential decision-making problem in which, at a given system state, a temperature threshold for the next \texttt{ON}/\texttt{OFF} switch of the heater is decided. Then, based on this threshold decision and the underlying governing equations, the system dynamics evolve until the indoor temperature reaches the threshold. At this point, a new threshold is chosen for the next switching and this sequence goes on and on as depicted schematically in Fig.\ref{Fig:2}. The control problem is to find the optimal sequence of thresholds that maximizes the average rewards accumulated in a long run, for a given reward (cost) function.\\

We cast the control problem as the five-tuple SMDP. The system state vector, $s$, is defined as $[T, T_o, h_s]^\top$. Temperature thresholds, $T^{\mathrm{th}}$, define the SMDP actions. The thresholds could be switch-\texttt{ON} ($T^{\mathrm{th}}_{\mathrm{ON}}$) or switch-\texttt{OFF} ($T^{\mathrm{th}}_{\mathrm{OFF}}$) thresholds. The decision epochs $\tau$'s are the timestamps when the state trajectory hits the switching manifolds or, in other words, when the indoor temperature reaches the temperature threshold. The transition probabilities and, in general, the semi-Markov kernel $p(s_{k+1}, t|s_k,a_k)$ depend on thermodynamics of the building (e.g., Eq.(\ref{Eq:dynamics})) and exogenous dynamics of the outdoor temperature. \textcolor{black} {We assume the building's model is unknown; therefore, the semi-Markov kernel is not accessible to the controller.}\\

Similar to the SMDP formulation, the threshold control policies, and hence, the actions ($T^{\mathrm{th}}$) could be either stochastic or deterministic. The policies and the actions are determined by the switching manifolds. That means a stochastic (or deterministic) policy is originated from a stochastic (or deterministic) switching manifold. Similarly, a parameterized switching manifold results in a parameterized policy. For stochastic threshold policies, we constrain the policy to Gaussian distributions of the form:
\begin{equation}
\pi^{\theta}(T^{\mathrm{th}}|s_k) = \frac{1}{\sigma_{\theta^{\sigma}}(s_k)\sqrt{2\pi}} \exp\left(-\frac{\left(T^{\mathrm{th}}-m_{\theta^{m}}(s_k)\right)^2}{2{\sigma_{\theta^{\sigma}}(s_k)}^2}\right),
\label{Eq:Gaussian}
\end{equation}
where, $m_{\theta^{m}}(s_k)$, and $\sigma_{\theta^{\sigma}}(s_k)$ are mean and standard deviation of the threshold temperature $T^{\mathrm{th}}$, that are parameterized by parameter vectors $\theta^{m}$ and $\theta^{\sigma}$, respectively ($\theta=[\theta^m, \theta^{\sigma}]^\top$). For deterministic policies the threshold temperatures are simply chosen as $T^{\mathrm{th}}=\mu^\theta(s_k)$.\\

{ \color{black}
It is worth noting that determining the optimal level of parameterization complexity, i.e., the number or the general shape of the switching manifolds, is not an easy task. With that said, the domain expertise could help with the choice of number and/or shape of the switching manifolds. In fact, one of the main reasons to use policy-gradient methods in this paper is that it enables incorporating the domain knowledge via choosing an appropriate class of policies. Choosing manifolds with fewer parameters expedites the learning process; however, it can compromise the control performance if the optimal control does not belong to the family of chosen manifolds.\\ 
}

The reward at a given state-action pair constitutes a discrete switching penalty and a continuous part penalizing energy consumption as well as temperature deviation from a desired temperature $T_d$:
\begin{equation}
r(s_k,a_k)=r_{sw}+\mathbb{E}\left[ \int_{\tau_k}^{\tau_{k+1}}r_e h_s(t)+r_c \left(T-T_d\right)^2 dt | S_k=s_k, A_k=a_k \right],
\label{Eq:reward2}
\end{equation}
where, $r_{sw}$, $r_e$, and $r_c$ were defined earlier in section \ref{systemdynamics}. Defining the transition rewards as such allows us to map the performance measure of the original control problem (Eq.(\ref{Eq:performance})) to the infinite-horizon average reward of the SMDP formulation (Eq.(\ref{Eq:SMDPAveRew1})) with one-to-one correspondence of $c(W_{t'},S_{N_{t'}},A_{N_{t'}})=r_e h_s(t)+r_c \left(T-T_d\right)^2$ and $g(S_k,A_k)=r_{sw}$. Having formulated the micro-climate control problem in the SMDP framework, \textcolor{black} {we next develop RL algorithms that can autonomously learn the optimal control policy with no knowledge of the system dynamics.}

\section{Reinforcement learning algorithm and implementation}
\label{RL}
 We can iteratively improve the control policy using the policy gradient formulas, i.e., Eqs.(\ref{Eq:gradstoch}) and (\ref{Eq:graddet}), if the performance measure and the differential value functions can be calculated at each iteration. \textcolor{black} {However, accurate calculation of these functions requires access to the underlying semi-Markov kernel or, equivalently, the system dynamics which are assumed unknown to the controller. Therefore, we resort to reinforcement learning techniques where sampled data are used to approximate the said functions.}\\
 
 We develop actor-critic RL algorithms in which the actor employs the average-reward gradient formulas to update and improve the policy parameters while the critic estimates the differential functions as well as the performance measure. For the critic estimation we use parameterized differential action-value $Q^w(s,a)$ and state-value $V^v(s)$ functions with parameter vectors $w$ and $v$, respectively. We employ temporal difference (TD) learning for the critic estimation of the differential value functions. We also use an estimation of the true average reward, denoted by $\bar{J}(\theta)$, that is learned via the same temporal difference error. We use the following TD errors ($\delta_k$) for the critic estimation of $V^v(s)$ and $Q^w(s,a)$, respectively: 
\begin{eqnarray}
\delta_k & = & r_k-\bar{J}_k \Delta \tau_k + V^{v_k}(s_{k+1})- V^{v_k}(s_{k}) \\
\delta_k & = & r_k-\bar{J}_k \Delta \tau_k + Q^{w_k}(s_{k+1},a_{k+1})- Q^{w_k}(s_{k},a_k),
\label{Eq:TD}
\end{eqnarray}
where, $\bar{J}_k$, $v_k$, and $w_k$ are the average reward and parameter vectors at epoch $\tau_k$. $r_k$ is the sample reward and $\Delta \tau_k=\tau_{k+1}-\tau_k$ is the sample transition time at epoch $\tau_k$. The average reward estimate is updated using the TD error as follows:
\begin{equation}
\bar{J}_{k+1} = \bar{J}_{k} + \alpha_{J} \frac{\delta_k}{\Delta \tau_k},
\label{Eq:update}
\end{equation}
where, $\alpha_{J}$ is the learning rate for the average reward update.\\

In view of the actor and critic updates, we now present pseudocodes for event-triggered control problem in continuing tasks with both stochastic and deterministic policies. Algorithm \ref{Algo:stochastic} shows the pseudocode for stochastic policies with eligibility traces while algorithm \ref{Algo:deterministic} shows its deterministic counterpart. Algorithm \ref{Algo:deterministic} is an event-triggered compatible off-policy deterministic actor-critic algorithm with a simple Q-learning critic (ET-COPDAC-Q). For this algorithm we use a compatible function approximator for the $Q^w(s_k,a_k)$ in the form of $(a_k-\mu^{\theta}(s_k))^\top{\nabla_{\theta}\mu^{\theta}(s_k)}^\top w + V^v(s_k)$. Here $V^v(s_k)$ is any differentiable baseline function independent of $a_k$, such as state-value function. We parameterize the baseline function linearly in its feature vector as $V^v(s_k)=v^\top \phi_v(s_k)$, where, $\phi_v(s_k)$ is a feature vector. To simplify the notation in algorithms \ref{Algo:stochastic} and \ref{Algo:deterministic} we drop the subscript $(.)_k$ and replace the subscript $(.)_{k+1}$ by a prime superscript $(.)'$; for instance, $s_k$ and $s_{k+1}$ are replaced by $s$ and $s'$. In the next section, we implement these algorithms on two different building models and assess their efficacy.\\

\begin{algorithm}
\DontPrintSemicolon
\SetAlgoNoLine
{Input:} a differentiable stochastic policy parameterization $\pi^{\theta}(a|s)$\;
{Input:} a differentiable state-value function parameterization $V^{v}(s)$\;
{Parameters}: $\lambda_v \in [0,1]$, $\lambda_{\theta} \in [0,1]$, $\alpha_v>0$, $\alpha_{\theta}>0$, $\alpha_{J}>0$ \;
Initialize $\bar{J} \in \mathbb{R}$ (e.g., to 0)\;
Initialize state-value and policy parameters $v \in \mathbb{R}^{d_v}$ and $\theta \in \mathbb{R}^{d_{\theta}}$ (e.g., to {0})\;
Initialize the state vector $s \in \mathcal{S}$\;
\BlankLine
 
$z_v \leftarrow 0$ ($d_v$-component eligibility trace vector)\;
$z_{\theta} \leftarrow 0$ ($d_{\theta}$-component eligibility trace vector)\;
 \Repeat(  forever \underline{\textit{when an event occurs}}){
 $a \sim \pi^{\theta}(.|s)$\;
 Execute action $a$ and \underline{\textit{wait till next event}}; then observe $s'$, $r$, $\Delta \tau$ \;
 $\delta \leftarrow r-\bar{J} \Delta \tau + V^{v}(s')- V^{v}(s)$\;
 $\bar{J} \leftarrow \bar{J}+\alpha_{J}\frac{\delta}{\Delta \tau}$\;
 $z_v \leftarrow \lambda_v z_v + \nabla_v V^v(s) $\;
 $z_{\theta} \leftarrow \lambda_{\theta} z_{\theta} + \nabla_{\theta}\log\pi^{\theta}(a|s)$\;
 $v \leftarrow v+\alpha_v \delta z_v$\;
 $\theta \leftarrow \theta+\alpha_{\theta} \delta z_{\theta}$\;
 $s \leftarrow s'$\;
 }

 \caption{Event-triggered actor-critic stochastic policy gradient for continuing tasks with variable-time intervals (with eligibility traces)}
 \label{Algo:stochastic}
\end{algorithm}

\begin{algorithm}
\DontPrintSemicolon
\SetAlgoNoLine
{Input:} a differentiable deterministic policy parameterization $\mu^{\theta}(s)$\;
{Input:} a differentiable state-value function parameterization $V^{v}(s)$ \;
{Input:} a differentiable action-value function parameterization $Q^{w}(a,s)$\;
{Parameters}: $\alpha_v>0$, $\alpha_w>0$, $\alpha_{\theta}>0$, $\alpha_{J}>0$ \;
Initialize $\bar{J} \in \mathbb{R}$ (e.g., to 0)\;
Initialize state-value, action-value, and policy parameters $v \in \mathbb{R}^{d_v}$, $w \in \mathbb{R}^{d_w}$ and $\theta \in \mathbb{R}^{d_{\theta}}$ (e.g., to {0})\;
Initialize the state vector $s \in \mathcal{S}$\;
Initialize a random process $\{\mathcal{F}\}$ for action exploration
\BlankLine
 
 \Repeat(  forever \underline{\textit{when an event occurs}}){
 $a = \mu^{\theta}(s)+\mathcal{F}$\;
 Execute action $a$ and \underline{\textit{wait till next event}}; then observe $s'$, $r$, $\Delta \tau$ \;
 $\delta \leftarrow r-\bar{J} \Delta \tau + Q^{w}(s',\mu^{\theta}(s'))- Q^{w}(s,a)$\;
 $\bar{J} \leftarrow \bar{J}+\alpha_{J}\frac{\delta}{\Delta \tau}$\;
 $v \leftarrow v+ \alpha_v \delta \nabla_v V^v(s)=v+ \alpha_v \delta \phi_v(s)$\;
 $w \leftarrow w+\alpha_w \delta \nabla_w Q^w(s,a)=w+\alpha_w \delta \left(a-\mu^{\theta}(s)\right)^\top{\nabla_{\theta}\mu^{\theta}(s)}^\top$\;
 $\theta \leftarrow \theta+\alpha_{\theta} \nabla_{\theta}\mu^{\theta}(s)\nabla_a Q^w(s,a)|_{a=\mu(s)}=\theta+\alpha_{\theta}\nabla_{\theta}\mu^{\theta}(s)\left({\nabla_{\theta}\mu^{\theta}(s)}^\top w\right)$\;
 $s \leftarrow s'$\;
 }

 \caption{Event-triggered COPDAC-Q for continuing tasks with variable-time intervals}
 \label{Algo:deterministic}
\end{algorithm}

\section{Simulations and results}
\label{results}
In this section we implement our proposed algorithms to control the heating system of a one-zone building in order to minimize energy consumption without jeopardizing the occupants' comfort. To this end, we first describe the building models that we use for simulations, followed up by designing the rewards to use for our learning-based control algorithms. Then, we explain the policy parameterization used in the simulations before we present the simulation results.

\subsection{Building models}
\label{BuildingModels}
We use two one-zone building models: a simplified linear model characterized by a first-order ordinary differential equation, and a more realistic building modeled in EnergyPlus software. The linear model for the one-zone building with the heating system is as follows:
\begin{equation}
C \frac{dT}{dt}+K(T-T_o)=u(t) \dot{Q}_h,
\label{Eq:linear}
\end{equation}
where, $C=2000 \, kJK^{-1}$ is the building's heat capacity, $K=325 WK^{-1}$ is the building's thermal conductance, and $\dot{Q}_h=13 \, kW$ is the heater's power. As defined earlier, $u(t) \in \{0,1\}$ is the control action defining the heater status, and $T_o=-10 \, \degree C$ is the outdoor temperature.\\

In addition to the simplified linear model, a more realistic building, modeled in EnergyPlus, is also used for implementing our proposed learning control algorithms. The building modeled in EnergyPlus is a single-floor rectangular building with dimensions of $15.240 \times 15.240 \times 4.572 \, m^3$ ($50 \times 50 \times 15 \, ft^3$). The walls and the roof are modeled massless with thermal resistance of $1.291\, m^2\,K/W$ and $2.456\, m^2\,K/W$, respectively. All the walls as well as the roof are exposed to the sun and wind, and have thermal and solar absorptance of 0.90 and 0.75, respectively. The floor is made up of a 4-inch h.w. concrete block with conductivity of $1.730\, W/m\,K$, density of $2242.585\, kg/m^3$, specific heat capacity of $836.800\, J/kg\,K$, and thermal and solar absorptance of 0.90 and 0.65, respectively. The building is oriented 30 degrees east of north. EnergyPlus Chicago Weather data (Chicago-OHare Intl AP 725300) is used for the simulation. An electric heater with nominal heating rate of $10\, kW$ is used for space heating. 
\subsection{Rewards}
Comfort and energy consumption are controlled by the rewards. Rewards in RL play the role of cost function in controls theory, and therefore, proper design of the rewards is of paramount importance in the problem formulation. The three-component reward design detailed in sections \ref{systemdynamics} and \ref{prelim} is adopted here with reward coefficients of $r_{sw}=-0.8 \, unit$, $r_e=-1.2/3600 \, {unit}\,s^{-1}$, and $r_c=-1.2/3600 \, unit\,K^{-2}\,s^{-1}$. Here, \textit{unit} is an arbitrary scale for quantifying the different reward components

\subsection{stochastic and deterministic policy parameterization}
We use Gaussian distributions for stochastic policies as described by Eq.(\ref{Eq:Gaussian}). For simplicity, we consider the case where the switch-\texttt{ON} ($T^\mathrm{th}_\mathrm{ON}$) and switch-\texttt{OFF} ($T^\mathrm{th}_\mathrm{OFF}$) thresholds are not functions of the states. Hence, either of these thresholds has a mean and a standard deviation parameter, constant in the state space. We can thus parameterize the mean and standard deviation vectors as:
\begin{eqnarray}
m_{\theta^{m}}(s_k) & = & {\theta^{m}}^\top \phi(s_k)  \nonumber \\
\sigma_{\theta^{\sigma}}(s_k) & = & \exp({\theta^{\sigma}}^\top \phi(s_k)),
\label{Eq:meanstandard}
\end{eqnarray}
where, $\theta^{m}=[\theta^m_{\mathrm{ON}}, \theta^m_{\mathrm{OFF}}]^\top$ and $\theta^{\sigma}=[\theta^{\sigma}_{\mathrm{ON}}, \theta^{\sigma}_{\mathrm{OFF}}]^\top$. We further assume $\theta^{\sigma}_{\mathrm{ON}} = \theta^{\sigma}_{\mathrm{OFF}}=\theta_{\sigma}$. $\phi(s_k)=[1-h_s, h_s]^\top$ is the state feature vector. We also approximate the state-value function as $V_v(s_k)=[v_1, v_2]^\top \phi(s_k)$.\\

In a similar fashion and assuming that switch-\texttt{ON} and -\texttt{OFF} thresholds are constant in the state space, we parameterize the deterministic policy in the form of:
\begin{equation}
     T^{\mathrm{th}} = \mu^{\theta}(s_k) = \theta^\top \phi(s_k),
     \label{Eq:deterpolicy}
 \end{equation}
where, $\theta=[\theta_{\mathrm{ON}}, \theta_{\mathrm{OFF}}]^\top$ is the policy parameter vector. We approximate the action-value function by a compatible function approximator as $Q^w(s_k,a_k)=(a_k-\mu^{\theta}(s_k))^\top{\nabla_{\theta}\mu^{\theta}(s_k)}^\top w + V^v(s_k)$ with $w=[w_1, w_2]^\top$. The state feature vector $\phi(s_k)$ and the state-value function $V^v(s_k)$ are defined the same as in the stochastic policy.
\subsection{Results}
Having set-up the simulation environment and parameterized the control policies and the related function approximators, we can now implement the learning algorithms \ref{Algo:stochastic} and \ref{Algo:deterministic}. In order to asses the efficacy of our learning-based control methods, we would better have the ground truth optimal switching thresholds to which the results of our learning algorithms should converge.\\

It is worth noting that, even with a simple and known model of the building, the posed micro-climate control problem does not fall into any of the classic optimal control categories, such as LQG or LQR. This is mainly because of the complex form of the performance metric $J$. With that said, we can still find the optimal controller by brute-force simulation. Within the class of threshold policies, for fixed outdoor temperature and no stochasticity in the system dynamics, the optimal control will constitute constant switch-\texttt{ON} and -\texttt{OFF} thresholds\footnote{In the case of the EnergyPlus model, the outdoor temperature is not constant, and hence, the constant-threshold policy will not be the optimal policy but it is very close to the optimal policy.}. This simple form of the optimal control policy makes the brute-force simulation computationally reasonable, and it also justifies our constant-threshold assumption in parameterizing the policies in the previous section.
\\

\begin{figure}
 \centering
 \includegraphics[width =1.0 \linewidth]{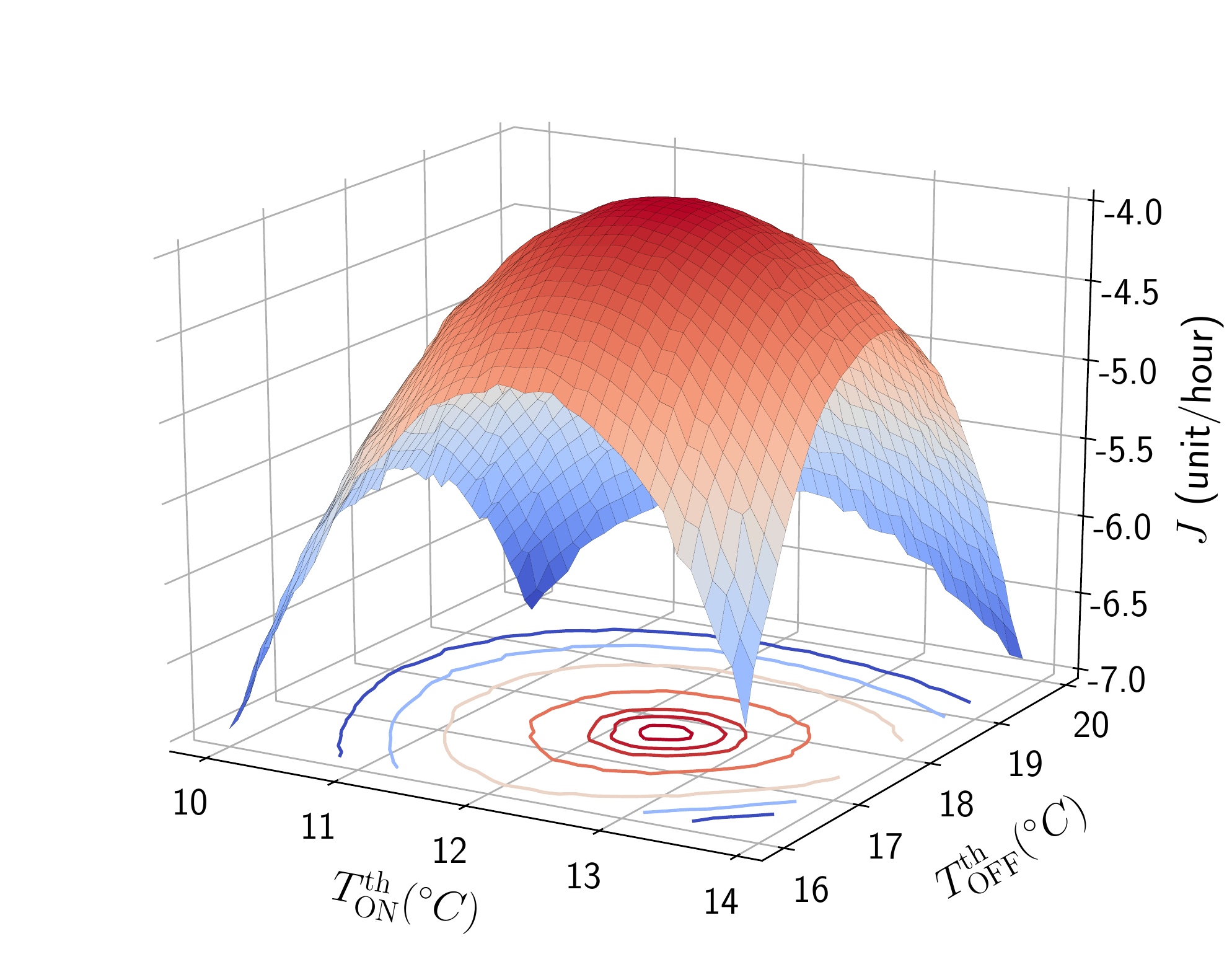}
 \caption{Average reward $J$ as a function of constant switch-\texttt{ON} ($T_{\mathrm{ON}}^{\mathrm{th}}$) and switch-\texttt{OFF} ($T_{\mathrm{OFF}}^{\mathrm{th}}$)  thresholds.}
  \label{Fig:3}
\end{figure}

To this end, we run numerous simulations where the system dynamics are described by either Eq.(\ref{Eq:linear}) or the EnergyPlus model, and the control policy is characterized by Eq.(\ref{Eq:deterpolicy}) with constant parameter vector $\theta$. For each such simulation, the simulation is run for a long time with a fixed pair of switching temperature thresholds. At the end of each simulation, the average reward is calculated by taking the ratio of the total accumulated reward to the total time. For the case where the system dynamics are described by Eq.(\ref{Eq:linear}), results are illustrated in Fig.\ref{Fig:3} based on which the optimal average reward is $J=-3.70 \, unit\,{hr}^{-1}$ corresponding to optimal thresholds of $T_{\mathrm{ON}}^{\mathrm{th}}=12.5 \, \degree C$ and $T_{\mathrm{OFF}}^{\mathrm{th}}=17.5 \, \degree C$. Knowing the ground truth optimal policy for the simplified linear model of the building, we next implement our proposed stochastic and deterministic learning algorithms on this building model. \\

\begin{figure}
 \centering
 \includegraphics[width =1.0 \linewidth]{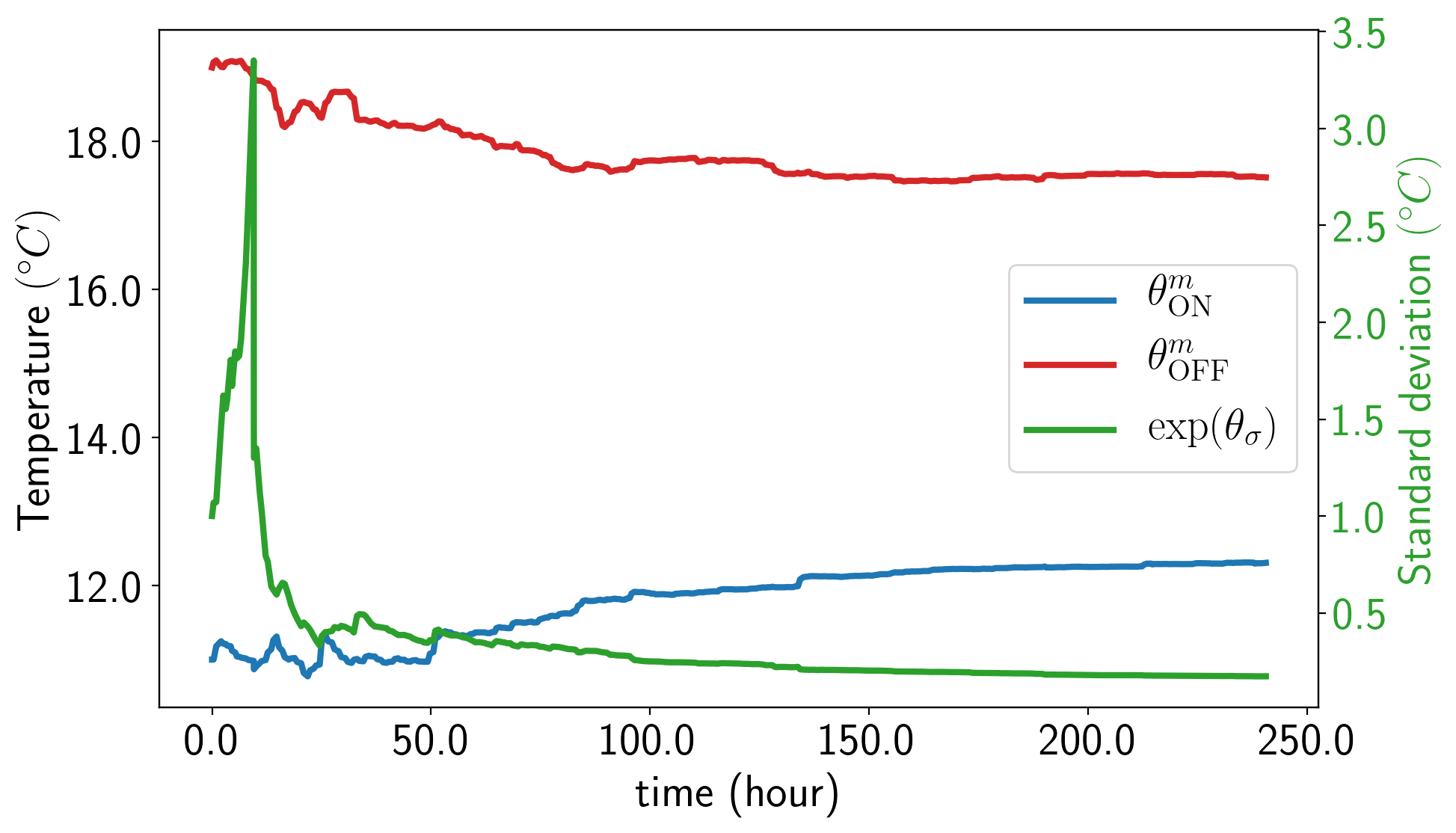}
 \caption{Time history of stochastic policy parameters, i.e., the means and standard deviation of the switching temperature thresholds, during a 10-day training by Algorithm \ref{Algo:stochastic}.}
  \label{Fig:4}
\end{figure}

\begin{figure}
 \centering
 \includegraphics[width =1.0 \linewidth]{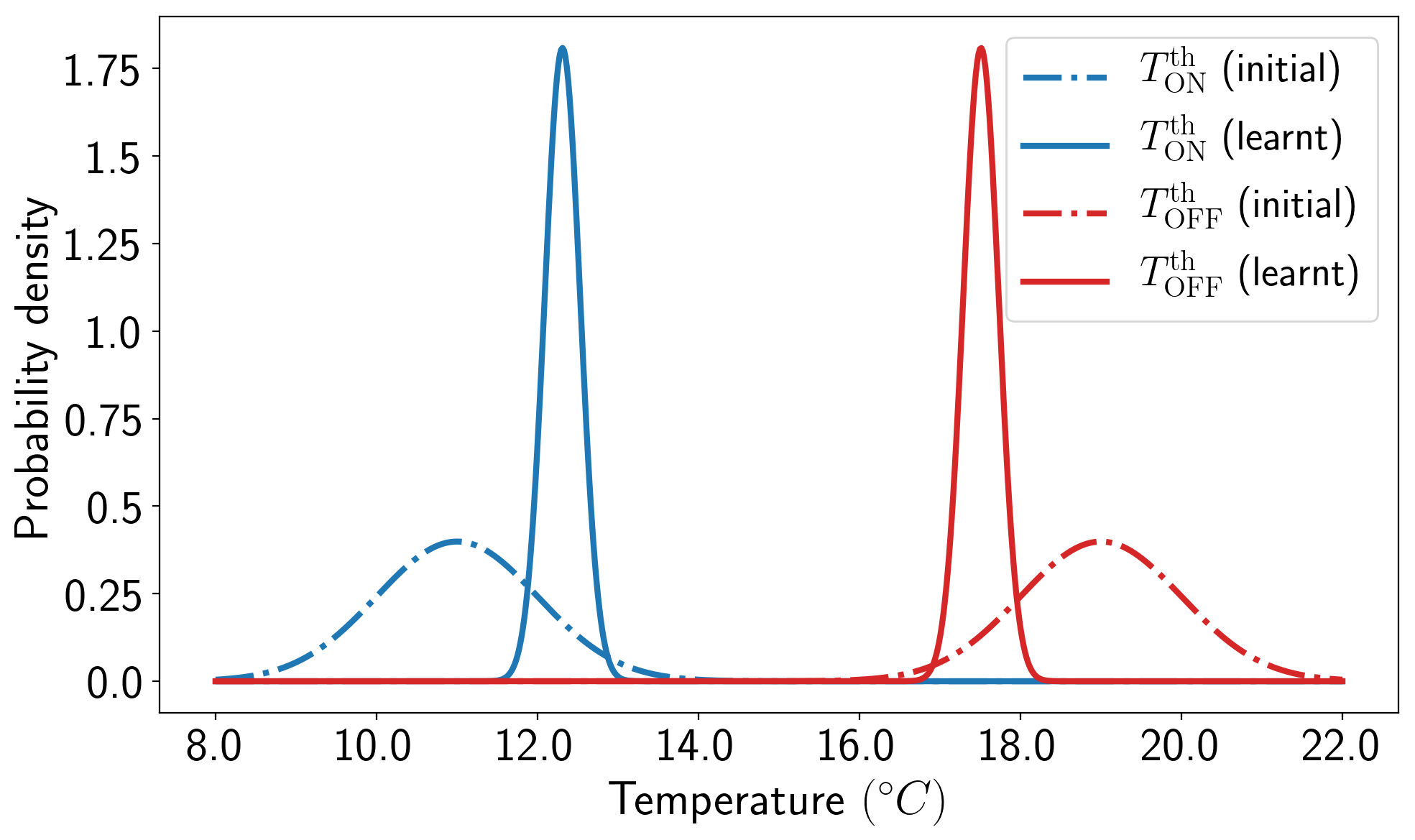}
 \caption{Initial and learned stochastic policies for switching temperature thresholds in a 10-day training by Algorithm \ref{Algo:stochastic}.}
  \label{Fig:5}
\end{figure}

\begin{figure}
 \centering
 \includegraphics[width =1.0 \linewidth]{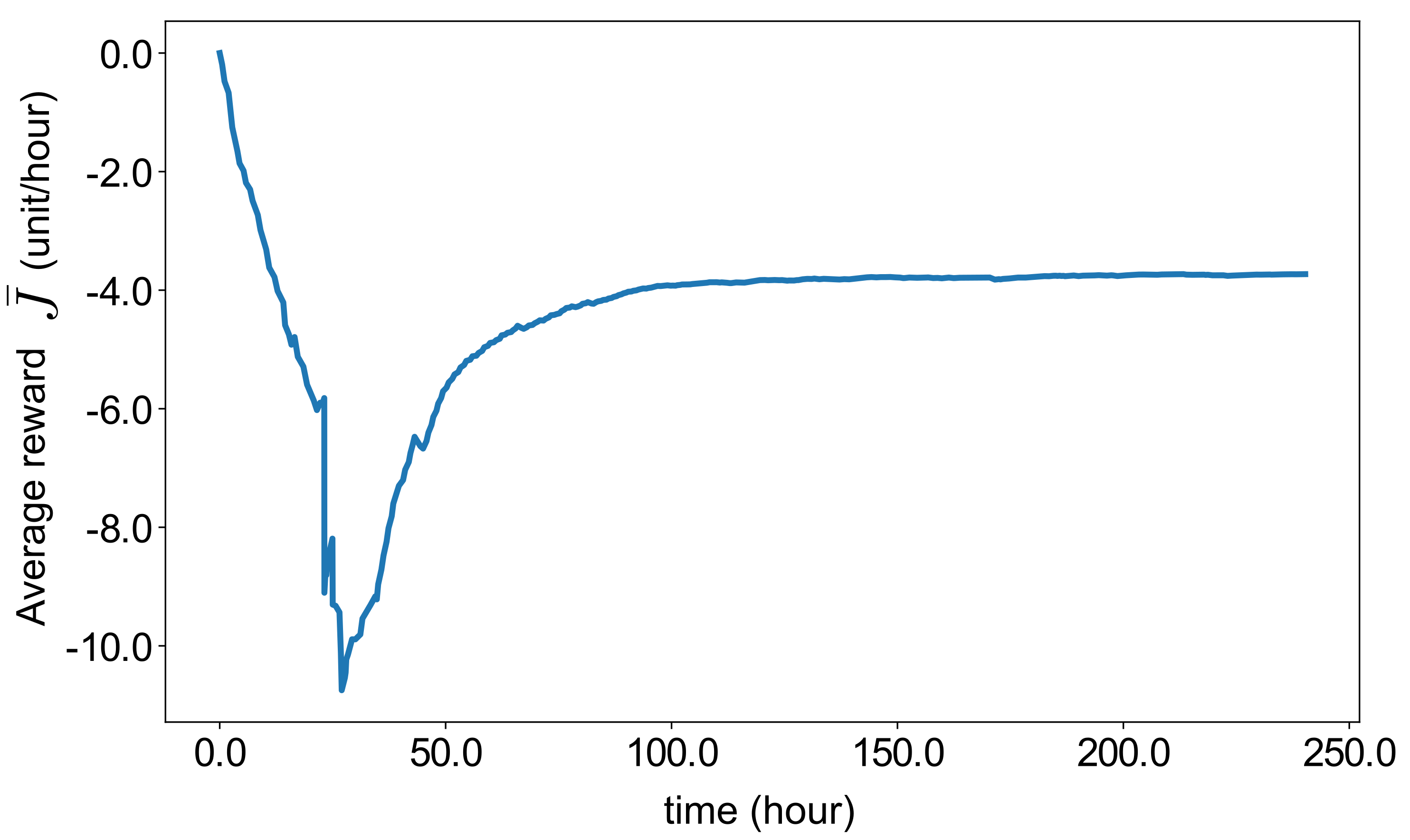}
 \caption{Time history of on-policy average reward in a 10-day training by Algorithm \ref{Algo:stochastic}.}
  \label{Fig:6}
\end{figure}

Figure \ref{Fig:4} depicts on-policy learning of the stochastic policy parameters during a training period of 10 days. Initial values of the mean of the threshold temperatures $[\theta^m_\mathrm{ON}, \theta^m_\mathrm{OFF}]$ are set to $[11.0, 19.0]\, \degree C$ and the initial standard deviation of these threshold temperatures are set to $1.0 \, \degree C$. Figure \ref{Fig:5} illustrates probability distributions of the stochastic policies for switching temperature thresholds before and after the 10-day training by Algorithm \ref{Algo:stochastic}. As seen in these two figures, the mean temperature thresholds have reached $12.3 \, \degree C$ and $ 17.5\, \degree C$, very close to the true optimal values. The standard deviation has decreased to $0.17 \, \degree C$ by the end of the training. According to Fig.\ref{Fig:6} the learned average reward converges to a value of $-3.73 \, unit\,{hr}^{-1}$. This learned policy is then implemented from the beginning in a separate 10-day simulation and the average reward is calculated as $-3.74 \, unit\,{hr}^{-1}$. Both of these values are very close to the optimal value of  $-3.70 \, unit\,{hr}^{-1}$, confirming the efficacy of the proposed event-triggered stochastic learning algorithm.\\

Next, we implement our deterministic event-triggered learning algorithm (Algorithm \ref{Algo:deterministic}) on the same building model. The learned \texttt{ON/OFF} switching temperature thresholds at the end of a 10-day training are found to be  $12.4 \, \degree C$ and $ 17.3\, \degree C$, again very close to the true optimal values. The implemented ET-COPDAC-Q is an \textit{off-policy} algorithm; hence, to assess its efficacy we need to calculate the \textit{on-policy} average reward of the learned policy. To this end, we implement the learned policy in a new simulation from the beginning and calculate the average reward at the end of the simulation. The on-policy average reward corresponding to the learned thresholds is then calculated to be $-3.73 \, unit\,{hr}^{-1}$ that is very close to the optimal value of $-3.70 \, unit\,{hr}^{-1}$.\\

In section \ref{limitations} we explained in detail that the traditional learning and control with fixed time intervals can deteriorate the learning quality or the control accuracy. We also claimed that our event-triggered learning control alleviate these drawbacks by exploiting the variable-interval mechanism for learning and control. To back this up via simulation, we run two 10-day simulations on the linear building model; one with our proposed variable-interval learning and control (Algorithm \ref{Algo:deterministic}) and the other with fixed intervals of 5-minute duration. The latter employs the same pseudocode as Algorithm \ref{Algo:deterministic} but applies learning updates and/or control executions at fixed time intervals. Our proposed approach learns the exact optimal thresholds, i.e., $12.5 \, \degree C$ and $ 17.5\, \degree C$ corresponding to an average reward of $-3.70 \, unit\,{hr}^{-1}$, whereas its fixed-interval counterpart learns the thresholds to be $11.2 \, \degree C$ and $ 19.3\, \degree C$.\\

\begin{figure}
 \centering
 \includegraphics[width =1.0 \linewidth]{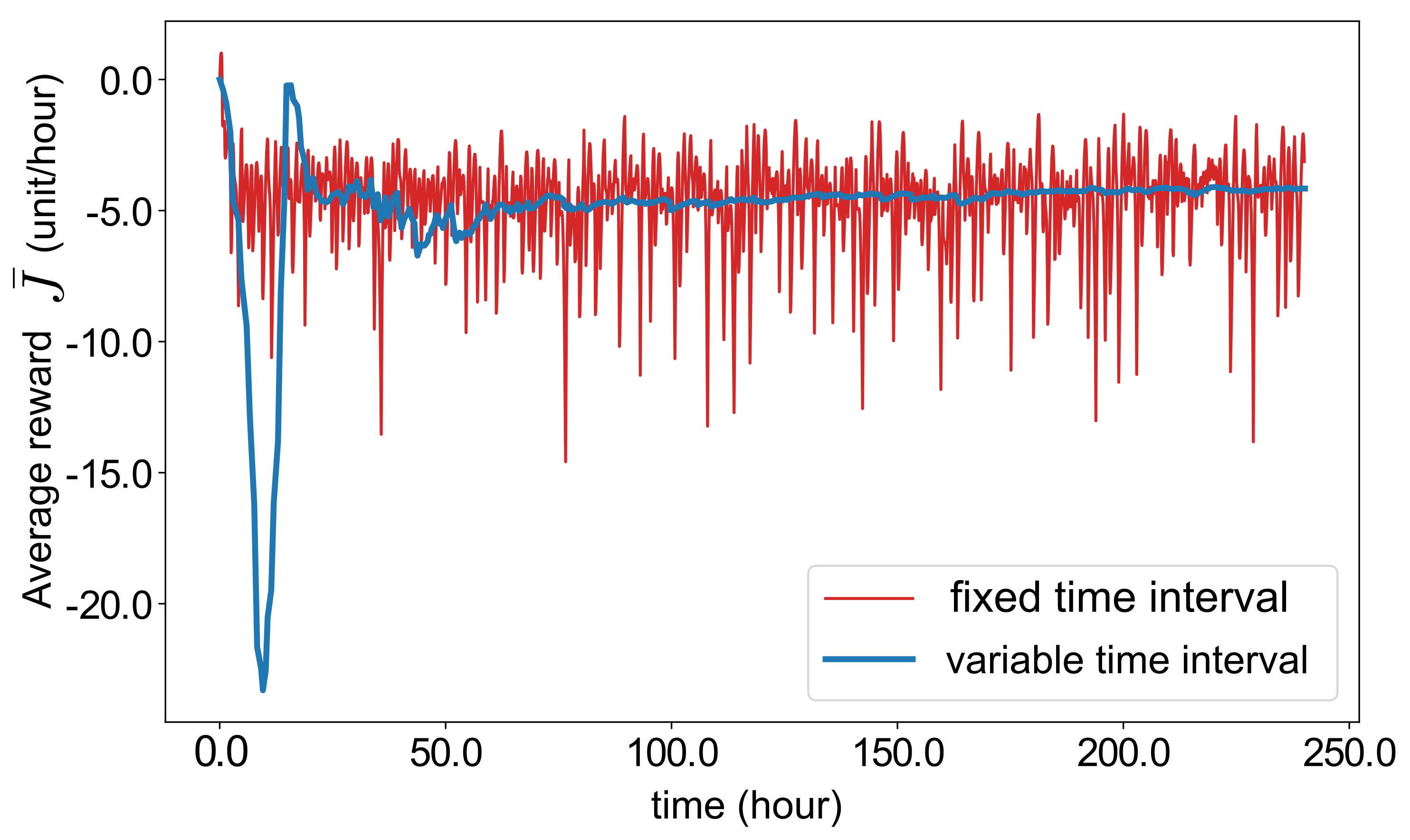}
 \caption{Time history of average reward in a 10-day training by Algorithm \ref{Algo:deterministic} with variable and fixed time intervals.}
  \label{Fig:7}
\end{figure}

If we now stop the training process (no learning updates), take the thresholds learned via the fixed-time interval approach, and implement them with control actions applied at fixed time intervals (i.e., both learning and control taking place with fixed time intervals), the average reward will be $-6.19 \, unit\,{hr}^{-1}$; however, this value improves to an average reward of $-5.22 \, unit\,{hr}^{-1}$, if the learned policy is implemented via event-triggered control (i.e., fixed time interval for learning but variable time interval for control). This corroborates the advantage of event-triggered learning and control over the classic time-triggered learning and control with fixed time intervals. To highlight this further, Fig.\ref{Fig:7} shows the learned average reward during a 10-day training by Algorithm \ref{Algo:deterministic} with both variable and fixed time intervals. It is clear that learning with fixed time intervals results in a considerably larger variance.\\

\begin{figure}
 \centering
 \includegraphics[width =1.0 \linewidth]{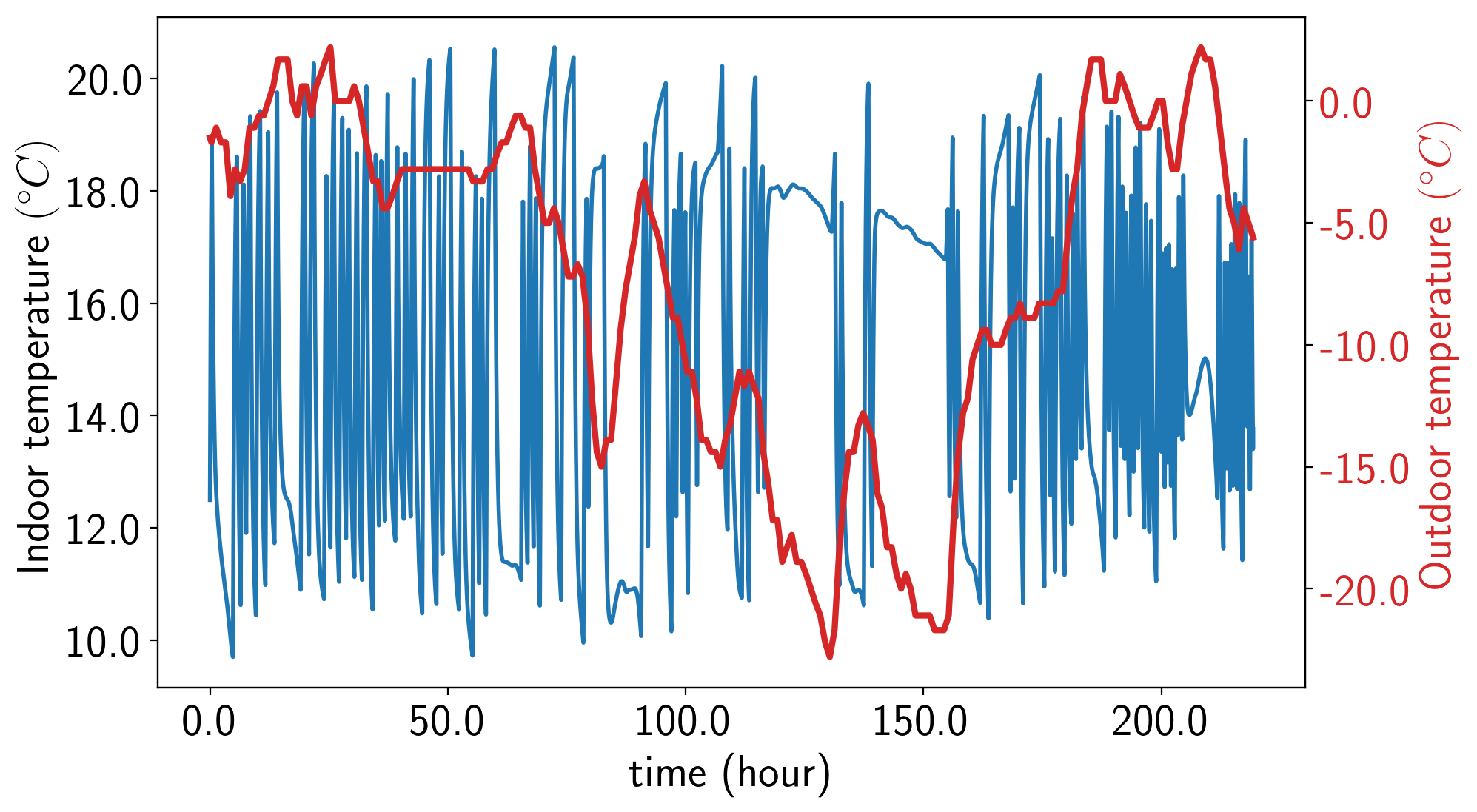}
 \caption{Time history of indoor and outdoor temperatures of the EnergyPlus building model during a 10-day training by Algorithm \ref{Algo:deterministic}.}
  \label{Fig:8}
\end{figure}

Last but not least, we implement our learning algorithms on the more realistic building, modeled in EnergyPlus software as detailed in section \ref{BuildingModels}. Here the outdoor temperature is no longer kept constant and varies as shown in Fig.\ref{Fig:8}. Although the optimal thresholds should, in general, be functions of outdoor temperature, here we constrain the learning problem to the family of threshold policies that are \textit{not} functions of the outdoor temperature. This is because (i) finding the ground truth optimal policy via brute-force simulations within this constrained family of policies is much easier than the unconstrained family of threshold policies, and (ii) based on our simulation results, the optimal policy has a weak dependence on the outdoor temperature in this setup.\\

\begin{figure}
 \centering
 \includegraphics[width =1.0 \linewidth]{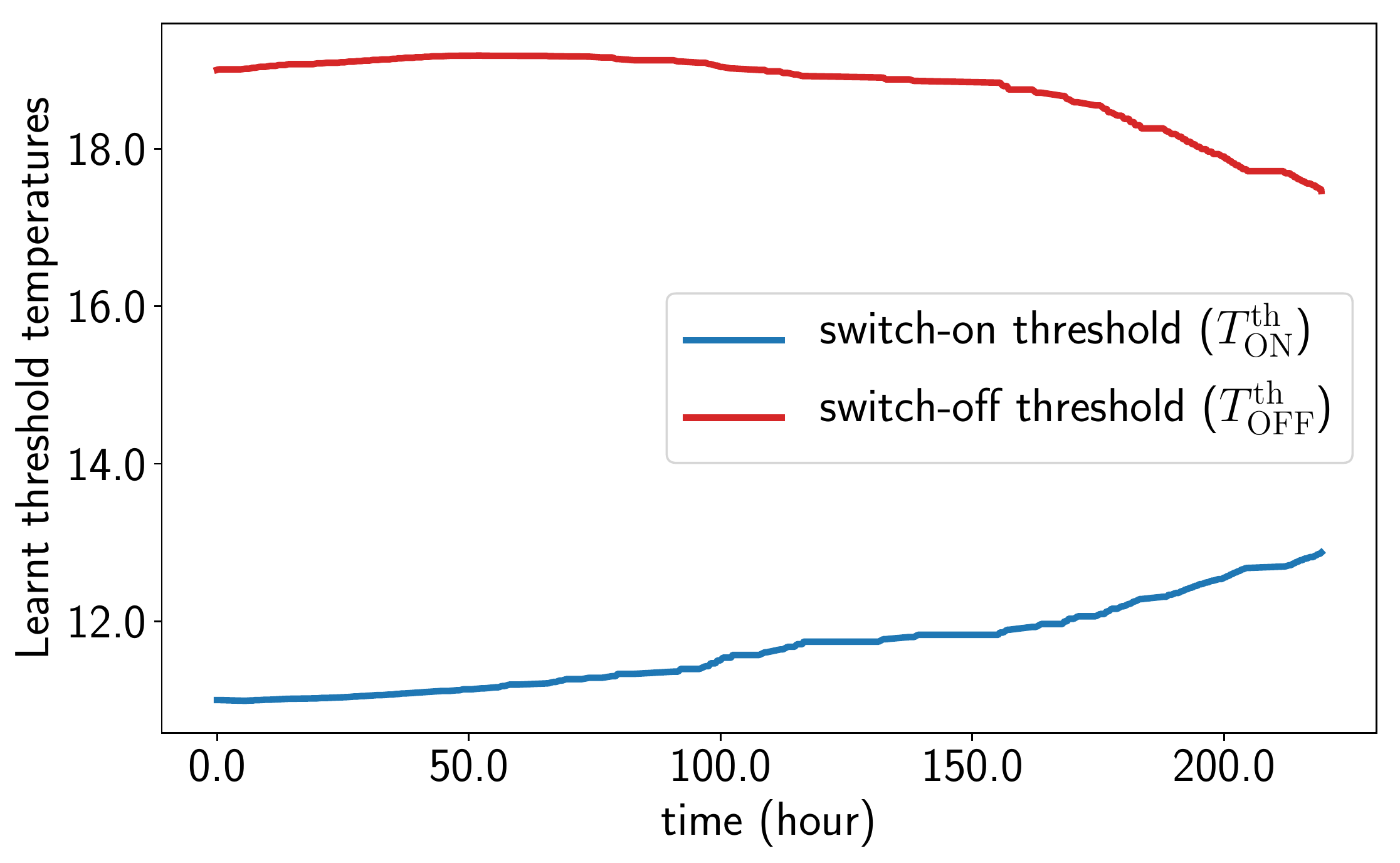}
 \caption{Time history of deterministic policy parameters, i.e., the switching temperature thresholds, during a 10-day training of the EnergyPlus building model by Algorithm \ref{Algo:deterministic}.}
  \label{Fig:9}
\end{figure}

Similar to the case of the simplified building model, we first find the optimal threshold policy and the corresponding optimal average reward by brute-force simulations. The optimal thresholds are found to be $T_{\mathrm{ON}}^{\mathrm{th}}=12.5 \, \degree C$ and $T_{\mathrm{OFF}}^{\mathrm{th}}=17.5 \, \degree C$ resulting in an optimal average reward of $J=-3.31\, unit\,{hr}^{-1}$. Here we employ our deterministic event-triggered COPDAC-Q algorithm to learn the optimal threshold policy. Starting from initial thresholds of $11.0 \, \degree C$ and $ 19.0\, \degree C$, the algorithm learns the threshold temperatures to be $12.9 \, \degree C$ and $ 17.5\, \degree C$ at the end of 10 days of training. This learned policy results in an average reward of  -$3.37\, unit\,{hr}^{-1}$. Time history of the building's indoor temperature controlled via an exploratory deterministic behaviour policy during the 10-day training period is illustrated in Fig.\ref{Fig:8}. Time history of the deterministic policy parameters, i.e., the switching temperature thresholds, during the 10-day training is shown in Fig.\ref{Fig:9}.

\section{Conclusion}
\label{conclusion}
This study focuses on event-triggered learning-based control in the context of cyber-physical systems with an application to buildings' micro-climate control. Often learning and control systems are designed based on sampling with \textit{fixed} time intervals. A shorter time interval usually leads to a more-precise controller but often degrades the learning performance by increasing the learning variance. To remedy these issues we proposed an event-triggered paradigm for learning and control with variable time intervals and showed its efficacy in designing a smart learning thermostat for autonomous micro-climate control in buildings.\\

We formulated the buildings' climate control problem based on a continuing-task SMDP with average reward setup. To reduce sample complexity of the learning-based control, we constrained the problem to the class of threshold control policies. The threshold policies are defined by their characteristic \textit{switching manifolds} in the state space. Control action switches only when the state trajectory hits one of these manifolds. Hitting the manifolds is referred to as \textit{events} -- hence the name \textit{event-triggered} control. The events trigger both learning and control processes.\\ 

 We employ policy gradient and temporal difference methods to learn the optimal switching manifolds that define the optimal control policy. Two event-triggered learning algorithms are proposed for stochastic and deterministic control policies. These algorithms are implemented on a single-zone building to concurrently decrease buildings' energy consumption and increase occupants' comfort. Two different building models are used: (i) a simplified model where the building's thermodynamics are characterized by a linear first-order ordinary differential equation, and (ii) a more realistic building, modeled in the EnergyPlus software. Simulation results show that the proposed algorithms learn the optimal policy in a reasonable one-week time. The results also confirm that, in terms of control performance and learning variance, our proposed event-triggered algorithms outperform their classic time-triggered reinforcement learning counterparts where both learning and control take place at fixed time intervals. The proposed algorithms improve the controller's performance measure by 70\%.\\

The variable-time flexibility of our proposed approach can benefit a wide range of control problems, such as coordination control in multi-agent systems. In many applications, proper coordination between different control agents is of paramount importance for achieving the global objective. For instance, coordination between different HVAC devices (e.g., heaters and ventilators) significantly affects the building's total energy consumption. Coordinated control of electric vehicles or thermostatically-controlled loads in a demand-response setup are two other coordination control problems that can benefit from our variable-time control framework. In addition, because of the SMDP formulation, our event-triggered control paradigm can be viewed as a hierarchical reinforcement learning, and hence, can potentially benefit from the recent advances in this field. \textcolor{black}{Last but not least, it is worth noting that despite all the advantages, the performance of the proposed learning-based controller is limited by the quality of the initially-chosen family of parameterized policies. Choosing a good policy class requires domain knowledge, and can become particularly hard with increased dimension of the states and actions.}

\section*{Acknowledgements}
This work is supported by the Skoltech NGP Program (joint Skoltech-MIT project).






\bibliographystyle{elsarticle-num}

\bibliography{sample}

\end{document}